\documentclass[12pt]{iopart}
\usepackage{epsfig}
\usepackage{amsfonts}
\usepackage{amssymb}
\begin{document}

\def\applss{\,\lower0.5ex\hbox{$\sim$}\kern-0.79em\raise0.5ex\hbox{$<$}\,}
\def\appgtr{\,\lower0.5ex\hbox{$\sim$}\kern-0.79em\raise0.5ex\hbox{$>$}\,}

\title[Chaotic angular-momentum pump]{Pumping angular momentum by
driven chaotic scattering}

\author{T Dittrich and F L Dubeibe}

\address{Departamento de F\'\i sica, Universidad Nacional de Colombia,
and \\
CeiBA -- Complejidad, Bogot\'a D.C., Colombia}
\ead{tdittrich@unal.edu.co}
\begin{abstract}
Chaotic scattering with an internal degree of freedom and the
possibility to generate directed transport of angular momentum is
studied in a specific model, a magnetic dipole moving in a
periodically modulated magnetic field confined to a compact region in
space. We show that this system is an irregular scatterer in large
parts of its parameter space. If in addition all spatio-temporal
symmetries are broken, directed transport of mass as well as angular
momentum occurs. The sensitive parameter dependence of the
corresponding currents includes frequent sign reversals. Zeros of
either quantity entail the exclusive occurrence of the other and thus
give rise in particular to angular-momentum separation without mass
transport as a classical analogue of spin-polarized currents.

\end{abstract}

\maketitle

\section{Introduction}\label{intro}
During two decades of research on chaotic scattering, internal degrees
of freedom have not enjoyed much interest. Yet there are numerous
reasons to give them a closer look: An internal freedom may provide
the additional dimension required to render a scattering system with a
single external degree of freedom chaotic
\cite{Jun87,Jun91,MB&95,PD&04}. It may also act as a reservoir that
absorbs energy from the external motion and stores it temporarily,
thus giving the scattering process a transient inelastic
character. Concerning applications, it is obvious that in chemical
reactions---one of the paradigms of chaotic scattering
\cite{NGR86,GR89}---the presence of internal freedoms is
indispensable. In mesoscopic physics, in turn, the basic entities are
electrons or fermionic quasi-particles: Their spin gives rise to a
rich phenomenology beyond point-particle physics and presently
receives much attention in the context of spintronics
\cite{Oes99,OY&99,KG02}.

In electronic and other mesoscopic systems, transport is an issue of
vital importance. It has been studied in depth in the framework of
adiabatic pumps \cite{Tho83,Bro98,AG99,JC&08}, electron counting
\cite{KJ&91,PL&92} and of the Kubo and B\"uttiker formulae
\cite{Lan57,Bue86,Bue88}. In spintronics, the practical necessity to
provide polarized currents has spurred research on directed spin
transport \cite{Oes99,OY&99,KG02}. In a different context, the study
of ratchets \cite{Rei00}, inspired by the biological phenomenon of
motor molecules \cite{Spu94,How01,Sch02}, has revealed general
physical conditions and numerous mechanisms for directed transport to
occur in systems far from thermal equilibrium. Also here, internal
freedoms are almost always involved and participate actively in
transport processes \cite{NK03}, e.g., as functional (configurational)
degrees of freedom in motor molecules \cite{MMK00,TS03} and as
internal heat baths.

In this paper we present a case study of chaotic scattering with an
internal degree of freedom, intrinsic angular momentum, and consider
how it contributes to irregular scattering and how in turn directed
transport of this quantity arises from scattering. The subject bears
on various of the research lines mentioned above: We shall invoke much
of what is known about chaotic scattering, in particular in
periodically driven systems \cite{HDR00,HDR01}. The coupling of a
magnetic dipole to an inhomogeneous magnetic field \cite{LW93} is
crucial for our modelling. Also concerning chaotic dynamics in such a
configuration we refer to earlier work \cite{Gar99}; in fact, it is
the essential ingredient of one of the first models proposed for
irregular scattering \cite{BS89}.

In directed transport, we take all the insights into account that
concern the importance of binary spatio-temporal symmetries and their
breaking \cite{FYZ00,DK&00,SO&01,SDK05}. We can here fall back on
direct precursor work on chaotic pumps \cite{DGS03}. The definition of
spin current is adopted from quantum contexts to define
angular-momentum transport in a classical setting. It should be
emphasized that we consider a driving that is both strong and
fast \cite{HDR00,HDR01} and thus not amenable to any perturbative
and/or adiabatic approximation. In this sense, we are dealing with
strongly nonlinear transport phenomena, far off the regime of linear
response and in particular of peristaltic pumping
\cite{Tho83,Bro98,AG99,JC&08}. As a consequence, directed transport is
achieved already driving the system via a \emph{single} parameter. The
crucial r\^ole of chaos also distinguishes our work from recent
studies of spin ratchets in ballistic electron systems
\cite{SW&06,SP&07,SB&07}. Preliminary results have been published on a
national platform \cite{DD08}.

As our main achievement, we present conclusive evidence that in
classical chaotic scattering systems, polarized currents can be
generated, i.e., directed transport of angular momentum occurs
independently of and even in the absence of mass (charge)
transport. We explain this effect in terms of the interplay of
asymmetry and the randomizing action of long unstable trajectories,
thus corroborating the decisive r\^ole of chaotic scattering. It is
also manifest in the sensitive parameter dependence of mass as well as
angular-momentum currents, which allows for a fine tuning of both
quantities. In particular, sign reversals are frequent and zeros of
one current give rise to the exclusive dominance of the other. In this
way, we achieve a ``rectification of angular momentum'' in the absence
of mass transport.

In \ref{model}, we construct our model and justify its
peculiarities. We discuss details of the dipole-field coupling that
have not been considered previously in the context of chaotic
scattering. Section \ref{chaoscat} is dedicated to irregular
scattering as the basic dynamical category of the system. Various
diagnostics are presented such as deflection functions, unstable
periodic orbits, and time-delay statistics. Our central subject,
directed transport, is approached in \ref{dirtrans}. We define the
relevant transport quantities in terms of reflection and transmission
coefficients. Numerical evidence is provided for directed mass as well
as for angular-momentum transport, and we point out the manifolds in
parameter space where pure currents of either kind occur. Section
\ref{conc} contains a summary and an outlook towards the quantization
of our model with the perspective to construct a chaotic spin
pump---evidently a major motivation for the present purely classical
study.

\section{A model for chaotic scattering with intrinsic angular
momentum}\label{model}

\subsection{Basic setup---coupling internal to external freedoms}
\label{setup}

In constructing our model, we attempt to satisfy the following
requirements, that (\emph{i}) the system show chaotic
scattering at least in parts of its phase space, as a dynamical
mechanism for directed transport, (\emph{ii}) the potential couple the
internal to the external degree of freedom and (\emph{iii}) vanish
identically outside a compact scattering region so that numerical
simulations can be restricted to a spatial box of finite extension,
and (\emph{iv}) be periodically time dependent in the form of delta
kicks in order to facilitate reducing the dynamics to discrete time,
that is, a map.

Moreover, previous work on nonlinear transport mechanisms in ratchets
and pumps has revealed a fundamental condition for directed currents
to occur \cite{FYZ00,DK&00,SO&01,SDK05,DGS03}: an inhomogeneous phase
space reflecting the absence of any spatio-temporal symmetry that
would give rise to pairs of otherwise identical trajectories carrying
the same current in opposite directions. It can be accomplished in
particular if the dynamic is mixed (regular coexisting with chaotic
motion) and the associated invariant manifolds exhibit some
asymmetry. We take these additional criterion into account
in setting up the model.

Specifically, we choose the internal freedom as an intrinsic angular
momentum (in the following we shall also refer to it as spin, wherever
its classical nature is obvious). In connection with a charge it gives
rise to a magnetic dipole moment that couples to an external magnetic
field and thus, if the field is not homogeneous, to the spatial motion
of the particle. However, we neglect the Lorentz force as concerns the
transversal components of the spatial motion. In this respect, we
treat the particle as neutral---in fact, neutrons would even provide a
more natural physical realization of our model than electrons.
Generally, this is justified as a good approximation if the
longitudinal velocity is sufficiently small. As an unwanted side
effect, with the Lorentz force we also loose the time arrow inherent
in the interaction with the magnetic field, as concerns orbital
motion. However, it remains effective through the spin dynamics, see
\ref{sym} below.

We can therefore restrict the set of relevant dynamical variables to
the longitudinal coordinate $x$ and its conjugate momentum $p$, as
well as the spin vector $\textbf{s} = (s_x,s_y,s_z)$. In terms of
these variables, the Hamiltonian reads
\begin{equation}\label{ham}
H(p,x,\textbf{s};t) = \frac{p^2}{2
m_0}-\gamma\,\textbf{B}(x,t)\cdot\textbf{s}\label{ham},
\end{equation}
where $m_0$ is the particle's mass and $\gamma$ the gyromagnetic
ratio.

We choose the space and time dependence of the field as
\begin{equation}\label{kicks}
\textbf{B}(x,t) = (0,B_1(x),B_2(x)) \sum_{n=-\infty}^{\infty}
\delta(n\tau-t),
\end{equation}
with
\begin{equation}\label{campo1}
B_1(x) = A_1 f(x+a/2), \quad B_2(x) = A_2 f(x-a/2),
\end{equation}
and an envelope function
\begin{equation}\label{enve}
f(x) = \exp\left[\frac{-1}{(a/2)^2 - x^2}\right]
 \Theta\left[a/2-|x|\right],
\end{equation}
as depicted in figure \ref{Bconfig}a. This function vanishes outside
the interval $[-a/2,a/2]$ yet is infinitely often differentiable.

\subsection{Equations of motion}

In the Hamiltonian (\ref{ham}), the spin $\textbf{s}$ plays an
exceptional r\^ole in that kinematically it should be treated as an
angular momentum, while there is no corresponding kinetic energy term,
and a canonically conjugate angle is not defined. For a correct
handling of these subtleties, we follow the development in
\cite{LW93} and obtain the equations of motion
\begin{equation}\label{eq2}
m_0 \ddot{\textbf{x}}=\gamma\nabla(\textbf{B}\cdot\textbf{s})\,,\quad
\dot{\textbf{s}}=\gamma\,\textbf{s}\times\textbf{B}.
\end{equation}
In all that follows, $m_0 = \gamma = \tau = 1$ is understood. An
important feature of (\ref{eq2}) is that the spin dynamics preserves the
form of a Lorentz force and thus breaks time-reversal invariance (TRI).

\begin{figure}[h!]
\begin{center}
\includegraphics[width=15cm]{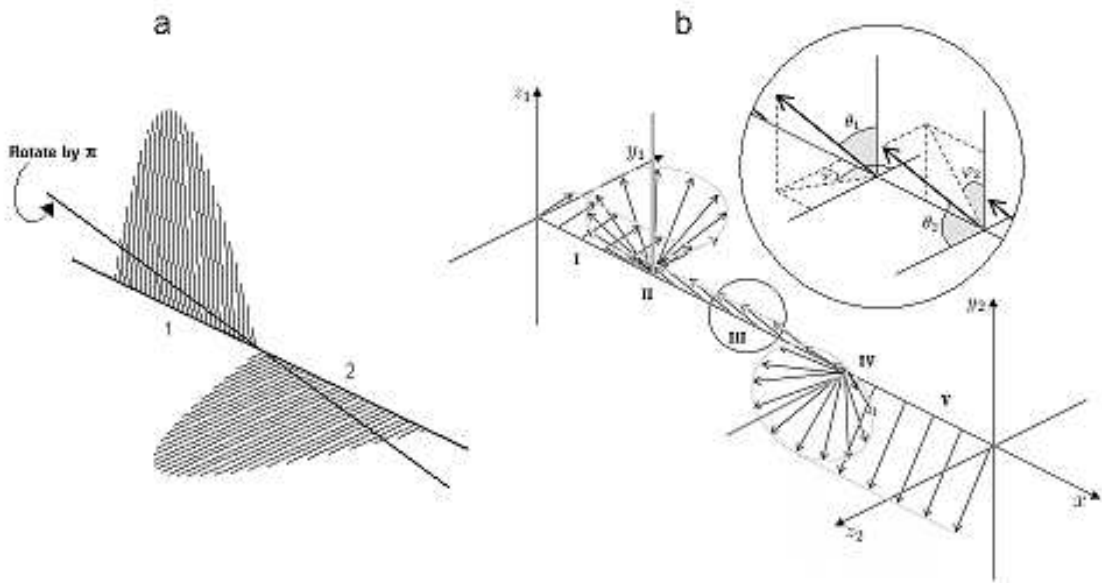}
\caption{Magnetic field and spin motion in the scattering region. (a)
Configuration of the magnetic field
(\protect\ref{campo1},\protect\ref{enve}). In each of the two sectors
1 and 2, the field is isotropic, with an angle of $\pi/2$ between the
two sectors. For equal amplitudes and identical envelopes, the field
is symmetric with respect to rotation by $\pi$ about the bold line,
corresponding to the transformation (\protect\ref{rotdiag}). Rendering
the widths and/or the amplitudes of the two field sectors different
breaks this symmetry. (b) Schematic representation of the scattering
process involving the spin vector $\textbf{s}$ (bold arrows). In the
incoming and outgoing asymptotic regions (phases I, V, resp.)
$\textbf{s}$ remains constant. Within the field sectors 1 and 2
(phases II and IV, resp.) it precesses around $\textbf{B}$. Upon
passing from sector 1 to 2 (phase III), the angles $\theta$, $\phi$
defining the orientation of $\textbf{s}$ undergo a passive
transformation, cf.\ (\protect\ref{campo}), as depicted in the
enlargement (inset).}
\label{Bconfig}
\end{center}
\end{figure}

The field configuration defined in (\ref{campo1},\ref{enve})
implies that within the intervals $[-a,0]$ and $[0,a]$, the field
varies only in magnitude but not in direction. This allows for further
simplifications. We identify the $y$-coordinate with the local field
direction, $\textbf{e}_y \equiv \textbf{B}/B$, $B \equiv
|\textbf{B}|$, choose spherical coordinates for the spin,
\begin{equation}\label{sspher}
s_x=|\textbf{s}|\sin\theta\sin\varphi, \quad
s_y=|\textbf{s}|\cos\theta,\quad
s_z=|\textbf{s}|\sin\theta\cos\varphi,
\end{equation}
and denote $s\equiv |\textbf{s}|=1$. Then the $y$-component of the spin,
$m\equiv s\,\cos\theta$, is the angular momentum canonically conjugate
to the azimuth $\varphi$, while the polar angle $\theta$ becomes a
constant of motion, so that the spin dynamics reduces to a mere
precession, see figure \ref{Bconfig}b.
\begin{equation}\label{ef1}
\ddot{x}= m\, \mathrm{d} B/\mathrm{d}x \qquad \dot{\varphi}= -B
\end{equation}
with $m={\rm{const}}$. Note that the sense of precession is inverted
under time reversal $t \to -t$ (keeping $B \to B$), revealing the
time arrow inherent in (\ref{eq2}).

\subsection{Symmetry considerations}\label{sym}

For an inhomogeneous but unidirectional magnetic field, according to
(\ref{ef1}), the projection of the spin onto the field direction
is a cyclic variable. This kind of dynamics therefore may give rise
to chaotic spatial motion, but will not affect the spin. In order to
induce ``spin flips'' (again abusing quantum terminology), we add a
second interaction sector where the field is perpendicular to that in
the first, as specified in (\ref{campo1}) (see figure
\ref{Bconfig}a). Upon passing from region 1 to 2 or back, the spherical
coordinates (\ref{sspher}) defining the spin orientation in the two
respective systems of reference aligned with the field direction in
either region undergo a passive transformation as follows, cf.\ the
inset in figure \ref{Bconfig}b,
\begin{equation}\label{campo}
\begin{array}{ll}
\theta_2  = \arccos(\sin\theta_1\cos\varphi_1),\qquad&
\theta_1  = \arccos(\sin\theta_2\sin\varphi_2),\qquad\\
\varphi_2 = \arctan(\cot\theta_1\csc\varphi_1),\qquad&
\varphi_1 = \arctan(\tan\theta_2\cos\varphi_2).\qquad
\end{array}
\end{equation}
If the widths and amplitudes of the two field sections are identical,
however, a spatial symmetry remains, namely rotation about the line
through the origin and diagonal between the two field directions, see
figure \ref{Bconfig}a. In Cartesian coordinates, it corresponds to the
operation
\begin{equation}\label{rotdiag}
\begin{array}{lll}
x' = -x,\qquad & y' = z,\qquad & z' = y.\qquad
\end{array}
\end{equation}
This symmetry would impede directed transport. In order to
break it, we choose the two amplitudes distinct, $A_1 \neq A_2$, which
gives the difference $A_2-A_1$ the meaning of a symmetry-breaking
parameter.

Besides spatial symmetries, time-reversal invariance deserves some
special consideration. As discussed in \cite{FYZ00,DK&00,SO&01,SDK05},
it must also be broken to achieve directed transport because otherwise
the existence of pairs of symmetry-related but counterpropagating
trajectories leads to an exact cancellation of currents. In
\cite{FYZ00,DK&00,SO&01,SDK05,DGS03}, this has been achieved by
imposing an asymmetric profile on the driving force. In our case, the
driving is symmetric under time reversal, while TRI is lifted by the
presence of a magnetic field. This effect, however, is not obtained
through the Lorentz force---which indeed we neglect, cf.\
\ref{setup}. Rather, it is the spin dynamics $\dot{\textbf{s}} =
\gamma\,\textbf{s}\times\textbf{B}$, cf.\ (\ref{eq2}), that is altered
under $t \to -t$.

\subsection{Discrete-time dynamics}

The impulsive driving allows us to integrate the equations of motion
from one kick to the next. Placing time sections immediately before
each kick, i.e., $t_n = n\tau - 0^+$, we arrive at the following
stroboscopic maps for the two respective field regions
\numparts
\begin{eqnarray}\label{map}
 p_{n+1} &=& p_{n }- 2 x_{i,n} B_i(x_{i,n}) \cos\theta_{i,n}/
 [(a/2)^2 -x_{i,n}^2]^{2}, \\
 \varphi_{i,n+1}&=& \varphi_{i,n}- B_i(x_{i,n})\,, \\
 x_{i,n+1}&=& x_{i,n}+ p_{n+1}\,,
\end{eqnarray}
\endnumparts
where $i = 1,2$, $B_i(x)$ refers to (\ref{campo1}), $\varphi_{i,n}$
refers to the frames related by (\ref{campo}), and $x_{i,n} = x_n -
(-)^i a/2$. One readily checks that these maps are canonical, that is,
the determinant of their stability matrix equals unity.

\begin{figure}[h!]
\begin{center}
\includegraphics[width=8cm]{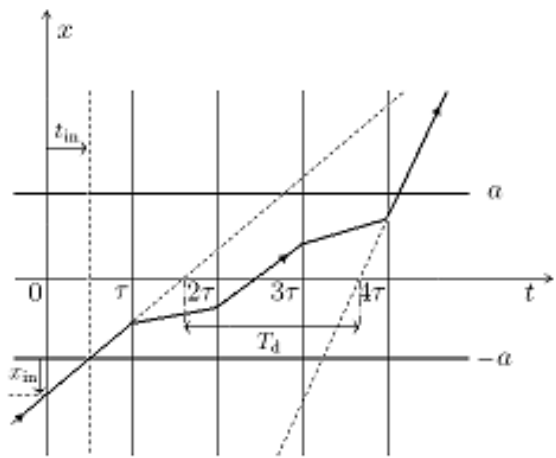}
\caption{The time lag $t_{\rm in}$ of a scattering trajectory (bold
zigzag curve) relative to the periodic kicks of the driving (vertical
lines) is defined with respect to the moment when it enters the
scattering region at $x = \pm a$. It can alternatively be scaled as a
phase shift $\phi_{\rm in} = 2\pi t_{\rm in}/\tau$ or a spatial offset
$x_{\rm in} = p_{\rm in}t_{\rm in}$. By contrast, the definition of
the delay time $T_{\rm d}$ refers to the extrapolation of the incoming
and outgoing asymptotes (dotted lines) forward and backward, resp.,
till they intersect the line $x = 0$. See text for details.}
\label{iniphase}
\end{center}
\end{figure}

In this periodically driven scattering system, the phase shift of an
incoming trajectory relative to the driving field appears as an
additional scattering parameter \cite{HDR00,HDR01,DGS03}. It can be
defined in various manners; for our particular case of an impulsive
driving and a compact spatial support of the field we choose the time
lag of the trajectory when it enters the field region at $x = \pm a$
with respect to the most recent kick at $t = n\tau$, cf.\ figure
\ref{iniphase}. By definition, its range is $0 \leq t_{\rm in} <
\tau$. Equivalently, it can be rescaled as a spatial shift $x_{\rm in}
= p_{\rm in}t_{\rm in}$, $0 \leq x_{\rm in} < p_{\rm in}\tau$, or a
phase angle $\phi_{\rm in} = 2\pi t_{\rm in}/\tau$, $0 \leq \phi_{\rm
in} < 2\pi$.

\subsection{Time scales}

The dynamics generated by (\ref{eq2}) is characterized by three
time scales: (\emph{i}) the temporal separation of the kicks, which
defines our unit of time, (\emph{ii}) the period of spin
precession, given by the inverse field strength $B^{-1}$, and
(\emph{iii}) the typical time to pass the scattering region, of the
order of $a/p_{\rm in}$. As it is our objective to study the
participation of the internal freedom in chaotic scattering, we shall
work in a regime where all the three time scales are comparable. This
prevents, in particular, using adiabatic approximations based on a
slow motion of the external coordinate as compared to spin precession.

By contrast, in figure \ref{Fig:fastpres}b below, we consider the case
of fast spin precession $A_i \gg 1$, $i = 1,2$, in order to
anticipate the regime of adiabatic spatial motion typical for
experiments with spin-$\frac{1}{2}$ systems.

\section{Chaotic scattering}\label{chaoscat}

We here consider chaotic scattering as defined by the following
properties \cite{Smi81,Ott93a,Ott93b}: (\emph{i}) the existence of a
chaotic repeller consisting of a discrete set of unstable periodic
orbits in the scattering region, (\emph{ii}) rapidly fluctuating
deflection functions with self-similar structure, at least in a
statistical sense, and singularities accumulating towards the orbits
of the chaotic repeller, and (\emph{iii}) an exponential distribution
of delay times inside the scattering region. In the subsequent
paragraphs we present numerical evidence that the system devised above
complies with all these criteria.

\subsection{Deflection functions}\label{deff}

\begin{figure}[h!]
\begin{center}
\includegraphics[width=14.5cm]{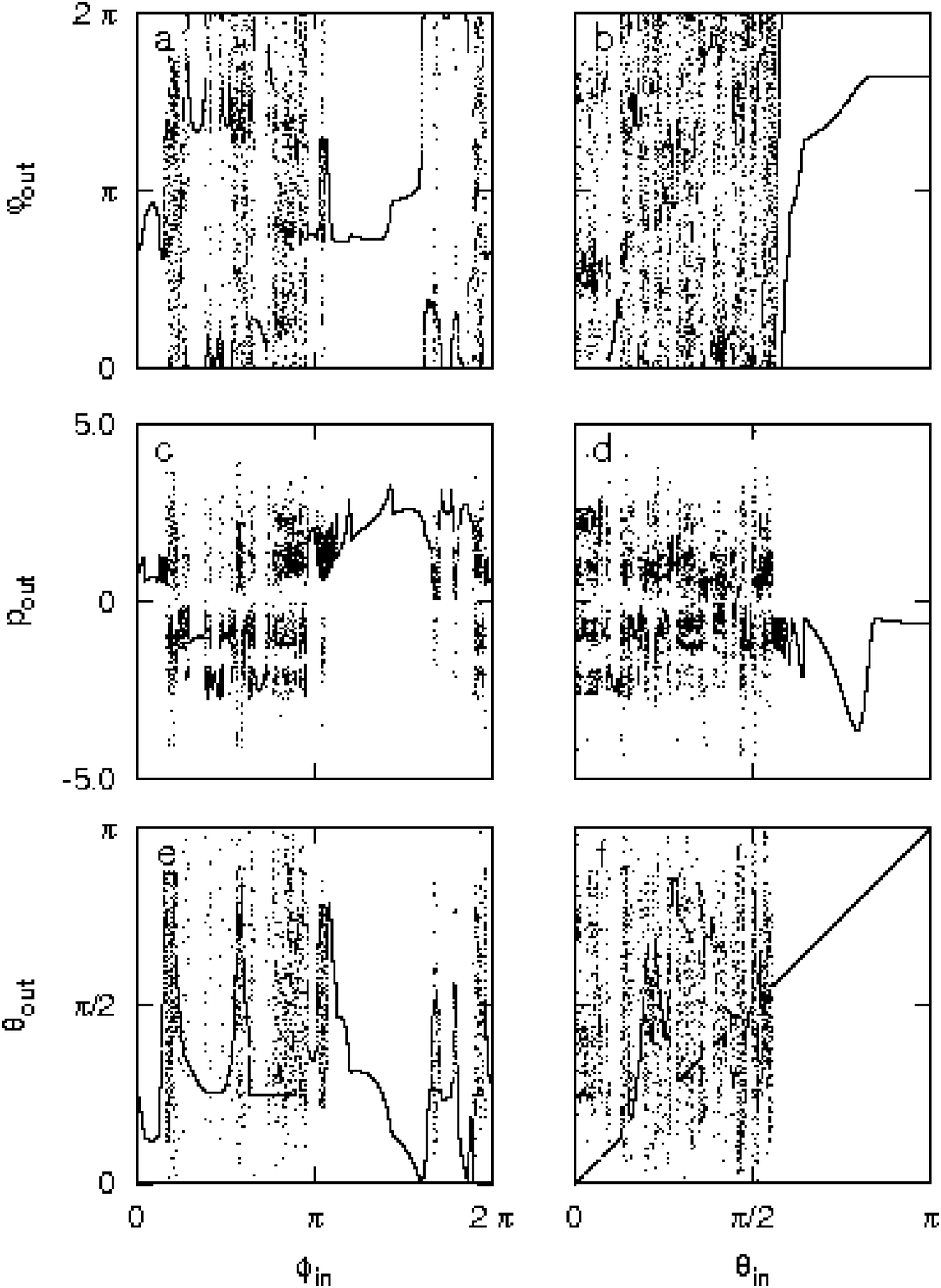}
\end{center}
\caption{Deflection functions. Outgoing azimuth
$\varphi_{\rm out}$ (a,b), outgoing linear momentum $p_{\rm{out}}$
(c,d), and outgoing polar angle $\theta_{\rm out}$ (e,f) vs.\
initial phase $\phi_{\rm in}$ (a,c,e) and vs.\ initial polar angle
$\theta_{\rm in}$ (b,d,f), respectively. The other initial
conditions and parameters are $p_{\rm in}=1$, $\theta_{\rm
in}=\pi/4$ (a,c,e), $p_{\rm in}=0.5$, $\phi_{\rm in}=0$ (b,d,f), and
$\varphi_{\rm in}=0$, $A_1=2$, $A_2=1$, $a=4$.} \label{Fig:disp}
\end{figure}

In plotting the deflection functions (figure \ref{Fig:disp}) we
concentrate on two parameters as incoming variables: the phase
$\phi_{\rm in}$ (left column, panels a,c,e) as an important control
parameter in experimental applications and the polar angle
$\theta_{\rm in}$ (right column, panels b,d,f) as we are interested
in the directed transport of angular momentum and ought to make sure
that this variable participates in the chaotic scattering.

While these figures show the typical behavior of a chaotic scatterer
for large parts of the parameter space, we observe a conspicuously
regular pattern in the right column of figure \ref{Fig:disp} (panels
b,d,f) for $\theta_{\rm in} \appgtr \pi/2$ that calls for a special
explanation. It reflects a strong asymmetry of the scattering
process: Trajectories entering region 1 from the left feel a strong
repulsive force and bounce back immediately without ever passing
into region 2, while trajectories coming in from the right undergo
the typical irregular scattering with unbounded delay time. Such long
trajectories tend to randomize the outgoing direction and thus lead to
approximately balanced transmission and reflection probabilities. We
shall see in section \ref{dirtrans} that this asymmetry is largely
responsible for the transport processes in the system.

\subsection{Stable and unstable periodic orbits}

\begin{figure}[h!]
\begin{center}
\includegraphics[width=5cm,angle=270]{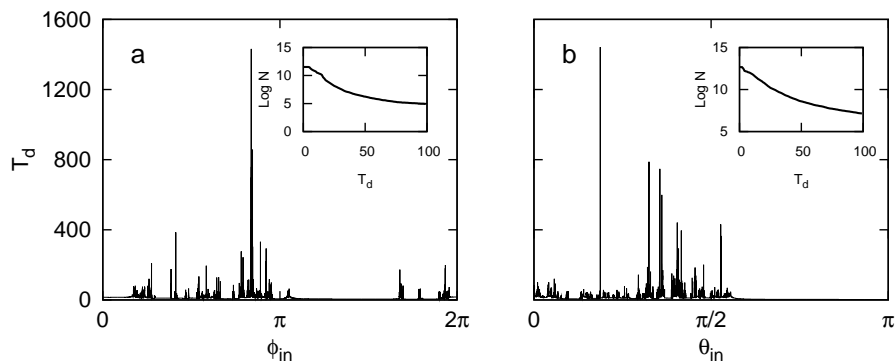}
\end{center}
\caption{Delay-time statistics as a function of initial
phase $\phi_{\rm in}$ (a) and incoming polar angle $\theta_{\rm in}$
(b). The other initial conditions and parameters are $p_{\rm
in}=1.0$, $\theta_{\rm in}=\pi/4$ (a), $p_{\rm in}=0.5$, $\phi_{\rm
in}=0$ (b), and $\varphi_{\rm in}=0$, $A_1=2$, $A_2=1$, $a=4$.}
\label{Fig:A12a4ft}
\end{figure}

From (\ref{map}) it is evident that the polar angle $\theta$ plays
a decisive r\^ole for the dynamics as its sign determines the local
stability of a trajectory. Specifically, upon passing from one field
region to the other, the dynamics turns elliptic or hyperbolic
according to whether $\cos \theta_{\rm in}$ is positive or negative,
respectively. This is clearly visible also in figure
\ref{Fig:A12a4ft}b as an absence of peaks for $\theta_{\rm in} >
\pi/2$.

\begin{figure}[h!]
\begin{center}
\includegraphics[width=5cm,angle=270]{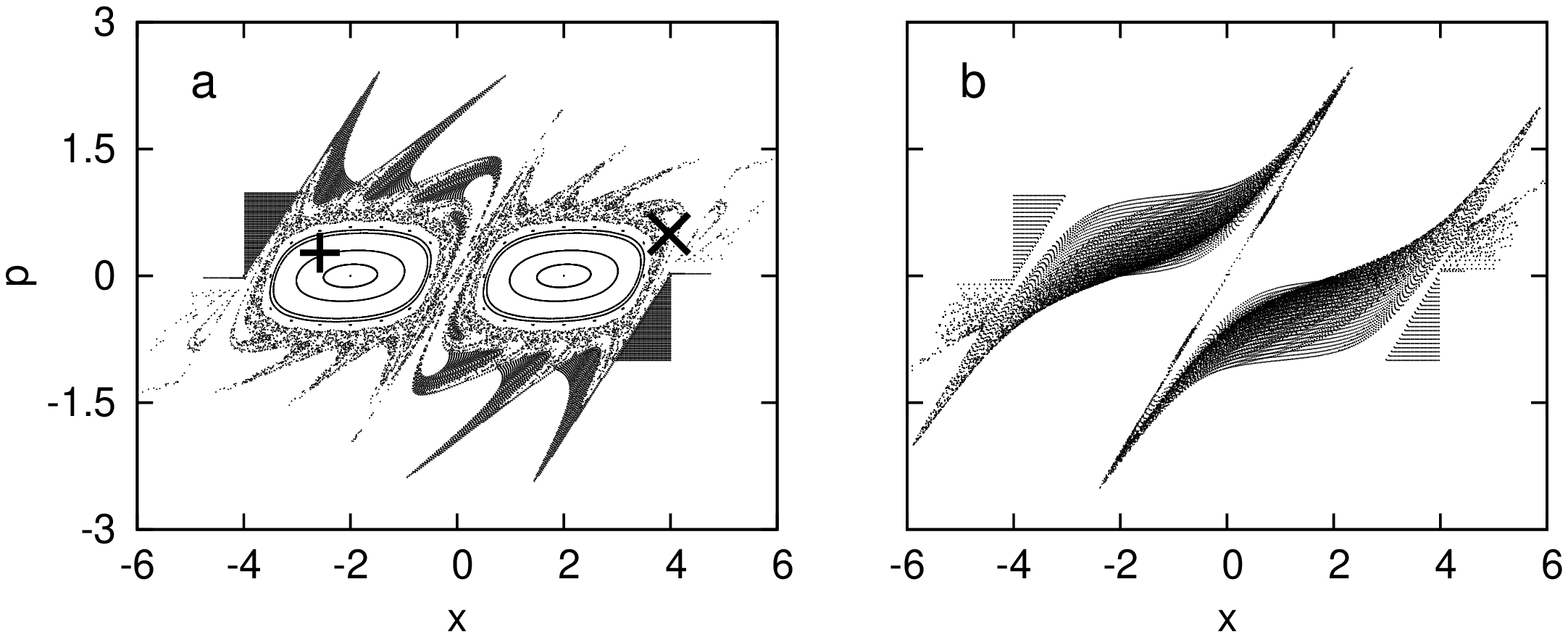}
\end{center}
\caption{Poincar\'e sections in the $(x,p)$-plane for
positive (panel a) vs.\ negative (b) incoming polar angle
$\theta_{\rm in}=-\pi/4$ (b). The other initial conditions and
parameters are $\phi_{\rm in}=0$, $\varphi_{\rm in}=0$, $A_1=A_2=1$
and $a=4$. The large regular areas in panel a correspond to stable
islands not accessible from outside the scattering region. Symbols
$+$ and $\times$ indicate periodic orbits depicted in figure
\protect\ref{Fig:po}a,b, resp.}
\label{Fig:sp}
\end{figure}

\begin{figure}[h!]
\begin{center}
\includegraphics[width=5.5cm,angle=270]{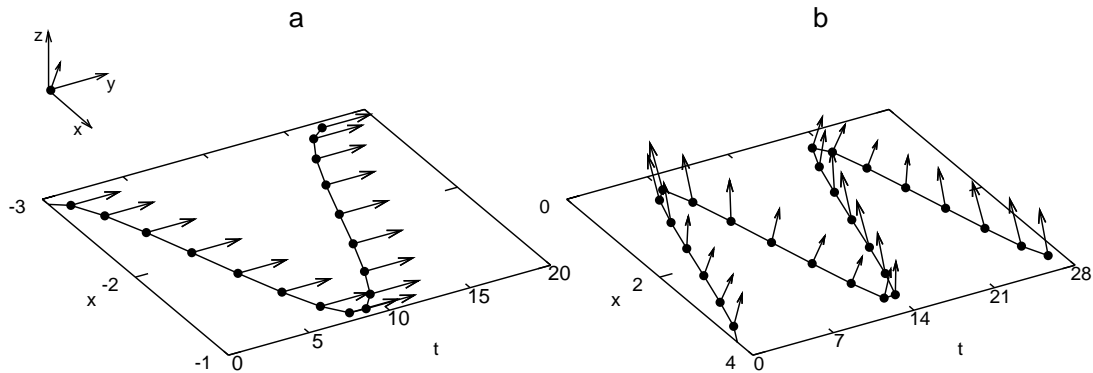}
\end{center}
\caption{Examples of a stable (a) and an unstable (b)
periodic orbit, located respectively in the left stable island in
figure \protect\ref{Fig:sp} (marked by $+$) and the chaotic region
($\times$). The spatial motion is shown as position $x$ vs.\
continuous time $t$, the spin orientation at each $t_n$ in an
independent 3-dim.\ coordinate system moving with the spatial
trajectory (inset). While the stable orbit leaves the spin at rest
parallel to the field, the unstable one involves a significant spin
motion.}
\label{Fig:po}
\end{figure}

In figure \ref{Fig:po} we depict a stable periodic orbit (a) that is
inaccessible from outside the scattering region and an unstable one
(b) that belongs to the chaotic repeller. They are marked by symbols
$+$ and $\times$, resp., in figure \ref{Fig:sp}a. While for the former,
the spin remains at rest, for the latter it exhibits a significant motion
that cannot even be reduced to mere precession. In orbital phase
space, this orbit forms a double loop, but the corresponding
self-crossing is lifted upon taking the spin motion into account.

\subsection{Delay-time statistics}

We define the delay time for a given scattering trajectory as usual as
the time difference between the incoming and outgoing asymptotes,
extrapolated forward and backward, respectively, till the origin, cf.\
figure \ref{iniphase}. This last specification is important as we are
dealing with a driven system where in general, energy is not conserved
and the outgoing momentum is different from the incoming one.

\section{Directed transport}\label{dirtrans}

We have shown in the preceding sections that our model fulfills the
two necessary conditions for directed transport mentioned above,
inhomogeneous phase space and absence of binary spatio-temporal
symmetries, and therefore expect to find directed currents at least in
parts of its parameter space.

\subsection{Asymmetric scattering and directed currents}

In the context of a scattering system, we define the current as the
frequency of particles leaving the scattering region to the right
minus the frequency of particles leaving to the left. Denoting
$T_{\alpha\beta}$ the fraction of particles transmitted from
``channel'' $\alpha$ to $\beta$ and $R_{\alpha\alpha}$ the fraction
of particles reflected from $\alpha$ back into $\alpha$, with
$\alpha,\beta = \textrm{``l''}$ (left) or $\textrm{``r''}$ (right),
the current is given by \cite{DGS03}
\begin{equation}\label{I}
I = \left(T_{\rm lr} + R_{\rm rr} - R_{\rm ll} - T_{\rm rl}\right).
\end{equation}

\begin{figure}[h!]
\begin{center}
\includegraphics[width=5.5cm,angle=270]{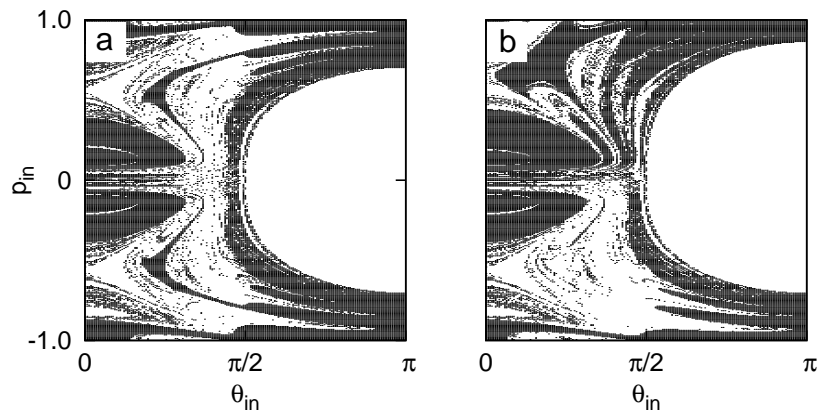}
\end{center}
\caption{Qualitative outcome of scattering, i.e., transmission
(black) vs.\ reflection (white) as a function of two initial
conditions, $\theta_{\rm in}$ and $p_{\rm in}$, for a configuration
with (a) and without (b) the spatial symmetry
(\protect\ref{rotdiag}). The other initial conditions and parameters
are $A_1 = 1$ (a), $A_1 = 2$ (b), and $A_2 = 1$, $a = 4$, $\phi_{\rm
in} = 0$, $\varphi_{\rm in} = 0$.} \label{Fig:trasim}
\end{figure}

\begin{figure}[h!]
\begin{center}
\hspace{0.3cm}\includegraphics[width=11cm]{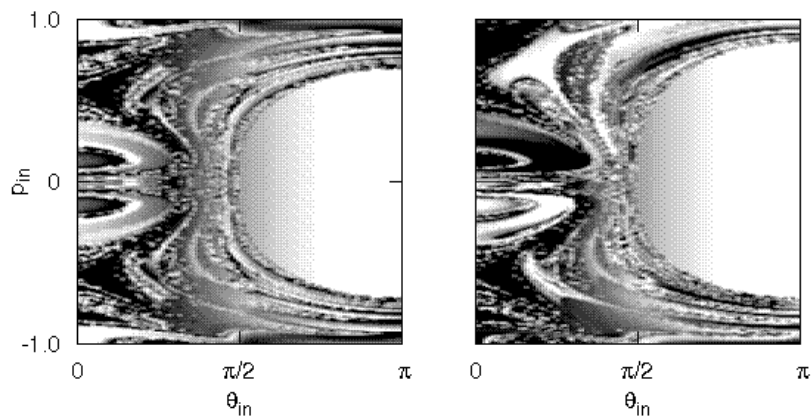}
\end{center}
\caption{Effective outgoing spin $\cos\theta_{\rm out}$ (white
$\equiv$ negative, grey $\equiv$ zero, black $\equiv$ positive) as a
function of incoming polar angle $\theta_{\rm in}$ and linear
momentum $p_{\rm in}$, for a configuration with (a) and without (b)
the spatial symmetry (\protect\ref{rotdiag}). The other initial
conditions and parameters are $A_1 = 1$ (a), $A_1 = 2$ (b), and $A_2
= 1$, $a = 4$, $\phi_{\rm in} = 0$, $\varphi_{\rm in} = 0$.}
\label{Fig:combisim}
\end{figure}

In the context of quantum electronics, the spin current is given
simply as the difference between the currents in the two spin states
``up'' vs.\ ``down'', $I_{\rm s} = I_{\uparrow}-I_{\downarrow}$, with
the partial currents defined in turn as in (\ref{I}). As an
immediate generalization to a larger total spin we would write the
spin current as a weighted sum of partial currents, with the weights
given by the projection of the spin onto some appropriate direction of
reference. This suggests to define the spin current in a classical
context, where the angular momentum is continuous, as a weighted
integral over partial currents parameterized by spherical coordinates,
\begin{equation}\label{Isi}
I_{\rm s} = \frac{1}{4\pi} \int_0^\pi {\rm d}\theta \sin\theta
\int_{-\pi}^\pi {\rm d}\varphi \cos\theta \, j(\theta,\varphi),
\end{equation}
with the current density
\begin{equation}\label{js}
j(\theta,\varphi) = \left(
j_{\rm lr}(\theta,\varphi) +
j_{\rm rr}(\theta,\varphi) -
j_{\rm ll}(\theta,\varphi) -
j_{\rm rl}(\theta,\varphi)\right),
\end{equation}
using obvious shorthands for the partial currents between the
respective leads. The angles should refer to some suitable laboratory
frame, here chosen as the system corresponding to field region 1, cf.\
(\ref{campo1}).

We here assume an initially unpolarized ensemble, corresponding to a
homogeneous angular-momentum distribution over the unit sphere. In
some cases it is preferable to fix part or all of the initial
conditions and to analyze the transport properties as a function of
these variables. In figures \ref{Fig:trasim} and \ref{Fig:combisim}, we
show respectively the outcome of the scattering process, i.e.,
transmission vs.\ reflection, and the polarization of the outgoing
particles as functions of the initial polar angle and linear
momentum. Also in these quantities we observe fractal self-similar
structures as in the deflection functions, and a correspondingly
sensitive parameter dependence. The presence (panels a) or absence
(panels b) of the symmetry (\ref{rotdiag}) is clearly reflected in the
transport features. The large almost void regions that appear in all
graphs for $\theta_{\rm in} > \pi/2$ correspond to trajectories that
bounce back immediately after entry, while manifestly asymmetric
structures are found only for $\theta_{\rm in} < \pi/2$. This confirms
the mechanism described in \ref{deff}, that chaotic randomization of
the outgoing direction occurs only for particles spinning in one sense
but not for the other.

The occurrence of directed transport in this system manifestly
violates linear-response theory. In terms of static quantities, we are
dealing with a non-zero mean (direct) current in the absence of any
external mean gradient. Even if we allow for time-dependent input and
output, the fact that the system is driven periodically through a
single parameter would entail an exact cancellation of transport in a
linear-response treatment. The same conclusion results from a Taylor
expansion of the current in terms of the driving frequency $\omega$,
\begin{equation}\label{lr}
I(\omega) = I_0 + I_1 \omega + O(\omega^2).
\end{equation}
If only the $I_1$-term were present (linear response), time reversal
of the driving $\omega \to -\omega$ would lead to an exact reversal of
the current (``turning the crank backwards pumps in the opposite
direction''). In our case, the driving is inherently symmetric in time
so that the presence of a directed current is incompatible with linear
response \cite{DGS03}. We emphasize that the functional dependence of
the current output on the driving amplitude is irrelevant for this
conclusion.

\begin{figure}[h!]
\begin{center}
\includegraphics[width=6cm,angle=270]{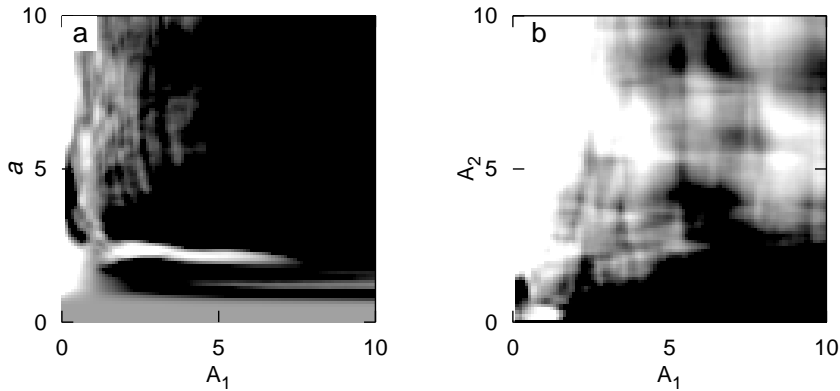}
\end{center}
\caption{Effective outgoing spin $\cos\theta_{\rm out}$ (white
$\equiv$ negative, grey $\equiv$ zero, black $\equiv$ positive) as
a function of width $a$ and amplitude $A_1$ of the left field
sector (a) and of the amplitudes $A_1$ and $A_2$ (b), averaged
over $\theta_{\rm in}$ and $p_{\rm in}$. The other initial
conditions and parameters are $A_2 = 1$ (a), $a = 4$ (b),
$\phi_{\rm in} =0$ and $\varphi_{\rm in} = 0$.}
\label{Fig:cspinaAA}
\end{figure}

The dependence of the polarization on width and amplitude of the
field, figure \ref{Fig:cspinaAA}a, demonstrates how the spin output
could be controlled varying these two parameters that are easily
accessible in experiments. The r\^ole of the symmetry is more
directly revealed in figure \ref{Fig:cspinaAA}b where the diagonal
$A_1 = A_2$ coincides with a zero of the polarization (corresponding
to a medium grey shade in the figure). There are, however, other
conspicuous zones of vanishing polarization apparently unrelated to
this symmetry.

\subsection{Angular-momentum transport}

\begin{figure}[h!]
\begin{center}
\includegraphics[width=11cm]{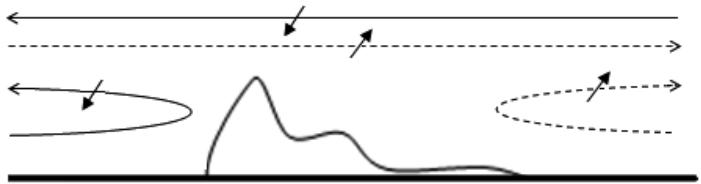}
\caption{A finite spin current in absence of net mass (charge)
transport can arise if particle currents cancel due to parity
invariance between transmission and reflection from either side,
while the corresponding coefficients for spin scattering lack this
symmetry. In this schematic figure, all particles with spin down
(full lines) leave to the left, irrespective of the incoming
direction, while those with spin up (dotted) all leave towards the
right. At the same time, if all currents are assumed to have the same
magnitude, mass transport vanishes identically, resulting in pure spin
separation.}
\label{Fig:polschem}
\end{center}
\end{figure}

\begin{figure}[h!]
\begin{center}
\begin{footnotesize}
\begin{tabular}{cc}
 \includegraphics[width=5.5cm,angle=270]{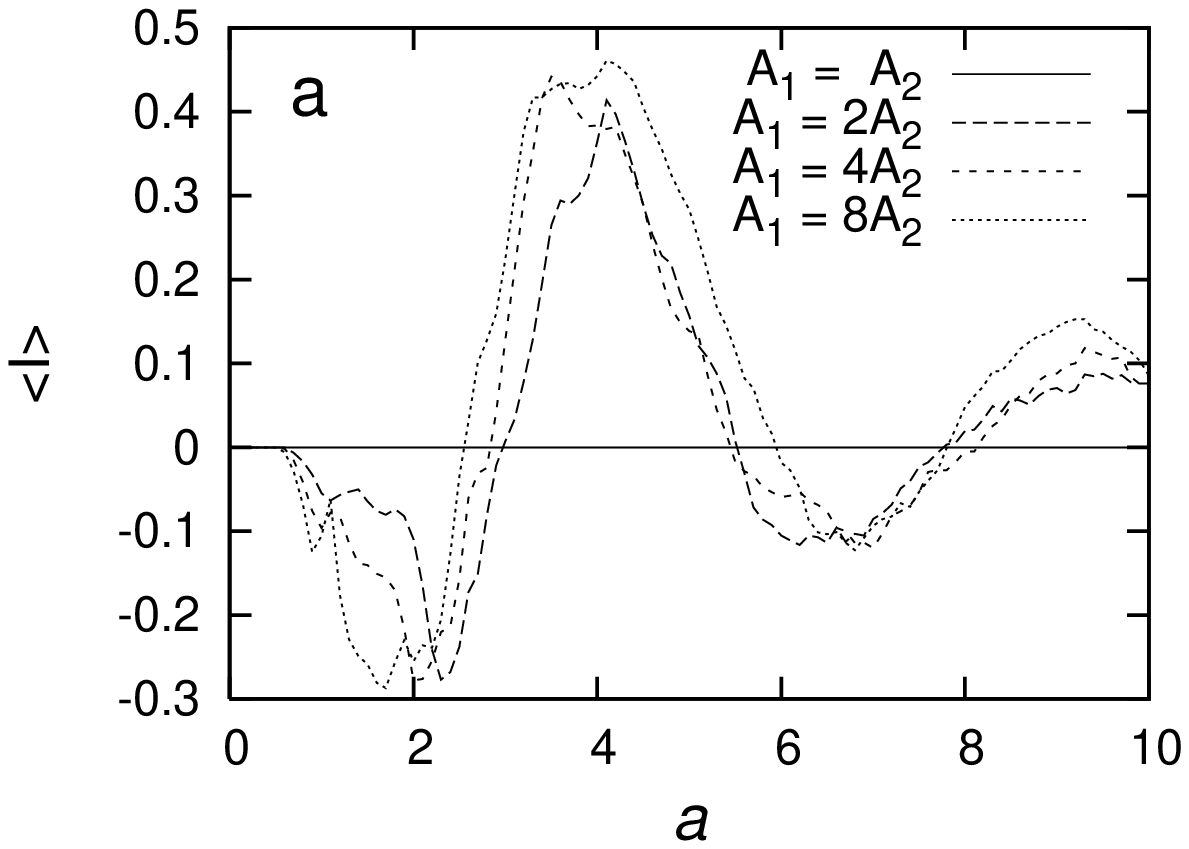}&
 \includegraphics[width=5.5cm,angle=270]{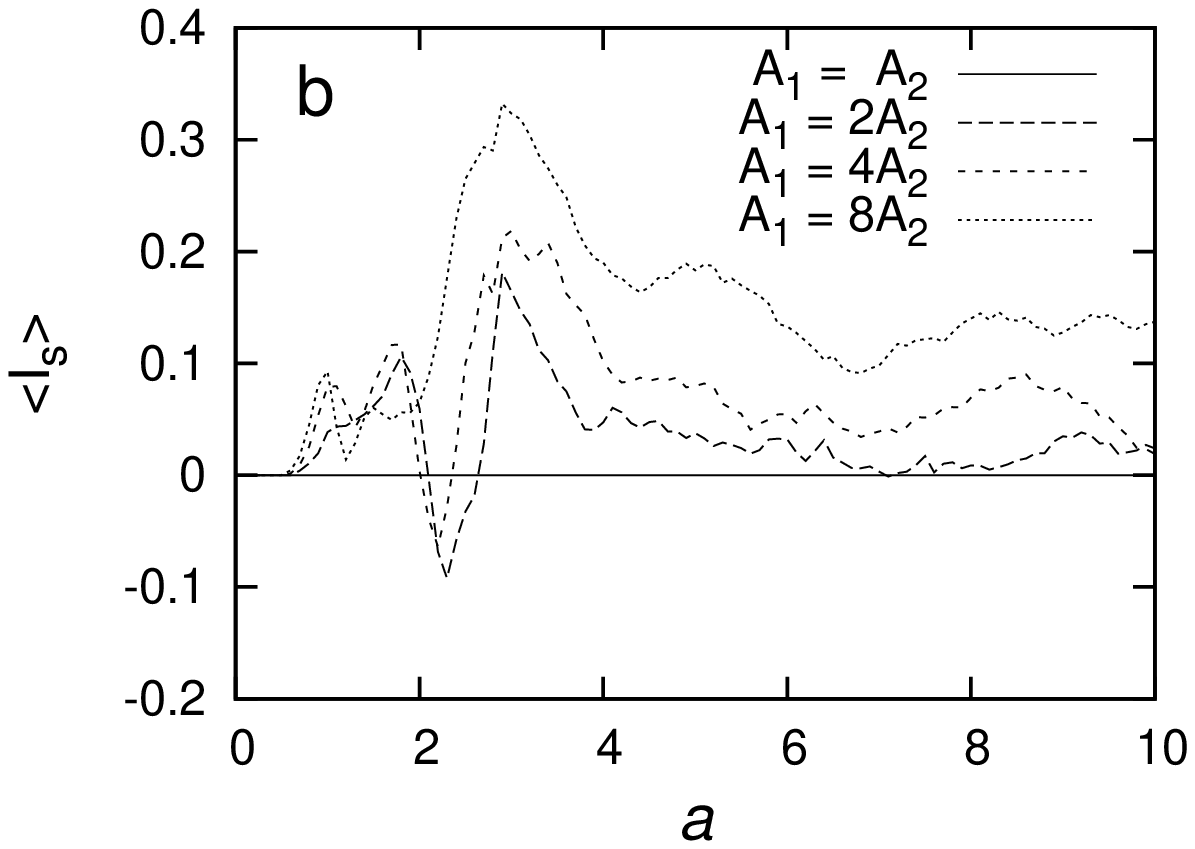}
\end{tabular}
\end{footnotesize}
\end{center}
\caption{Particle current $I$ (panel a) and spin current $I_{\rm s}$
(b), averaged over $\theta_{\rm in}$ and $p_{\rm in}$ vs.\ the width
of the well $a$ for various values of the ratio $A_1/A_2$.
Parameters are $A_2 = 1$ and $\varphi_{\rm{in}}=0$.}
\label{Fig:cpartaAA}
\end{figure}

\begin{figure}[h!]
\begin{center}
\begin{footnotesize}
\begin{tabular}{cc}
 \includegraphics[width=5.5cm,angle=270]{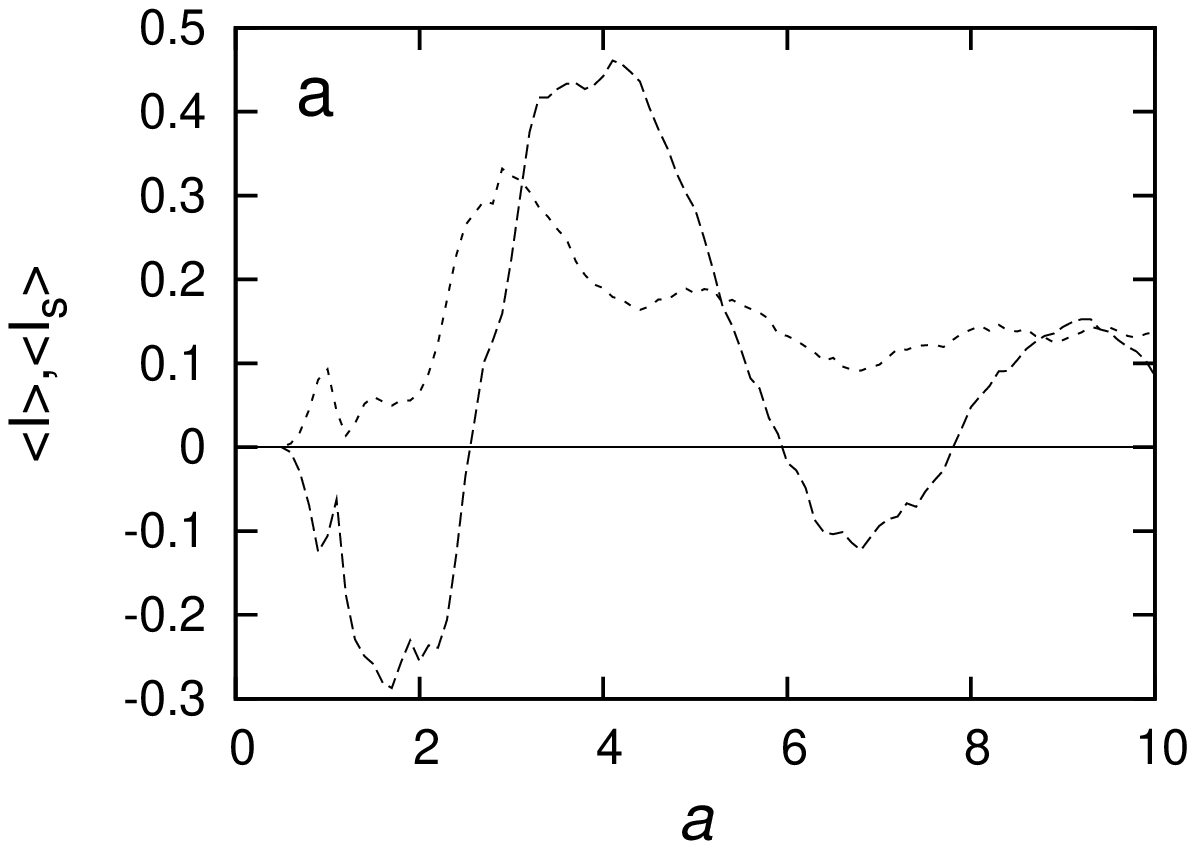}&
 \includegraphics[width=5.5cm,angle=270]{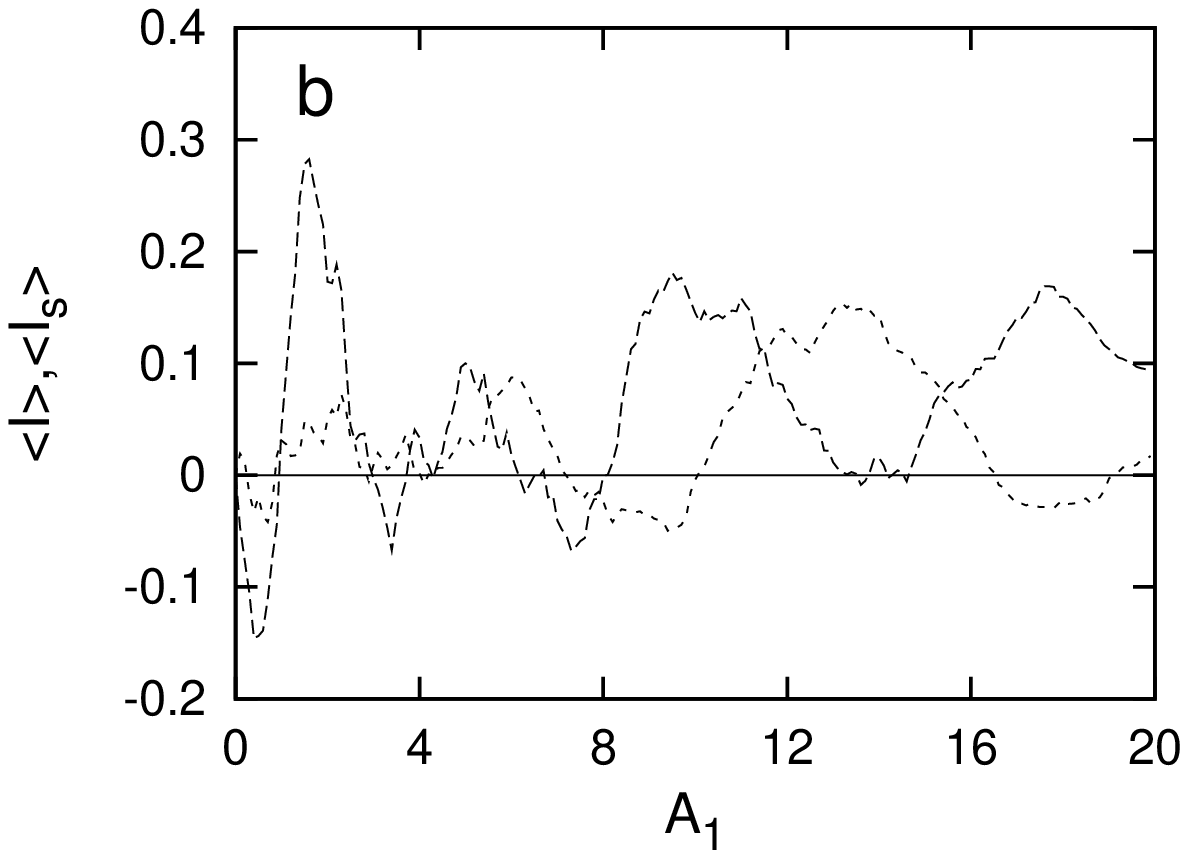}
\end{tabular}
\end{footnotesize}
\end{center}
\caption{Comparison of particle current $I$ (broken curve) to spin
current $I_{\rm s}$ (dotted), averaged over $\theta_{\rm in}$ and
$p_{\rm in}$, as functions of the width $a$ of the scattering region
(panel a) and the mean magnetic field strength (panel b). Parameters
are $\varphi_{\rm{in}}=0$ and $A_1 = 8$, $A_2 = 1$ (panel a) and $a =
4$, $A_1/A_2=1.5$ (b).} \label{Fig:fastpres}
\end{figure}

The data presented in figures \ref{Fig:combisim} and
\ref{Fig:cspinaAA} strongly indicate the existence of polarized
currents at least in certain regions of the parameter space. We
sketch in figure \ref{Fig:polschem} how a separation of spins could
come about in principle in the absence of a net mass current. In
order to corroborate its robustness and experimental feasibility, we
show in figures \ref{Fig:cpartaAA}, \ref{Fig:fastpres} global
averages both of the particle and the spin current, cf.\
(\ref{I},\ref{Isi}). We find zeros of the spin current $I_{\rm
s}$ at appreciable values of the particle current $I$, corresponding
to unpolarized charge transport, as well as zeros of $I$ at high
values of $I_{\rm s}$. These latter cases amount to a separation of
different orientations of angular momentum without net particle
transport, that is, to the classical analogue of spin filtering or
rectification.

Moreover, in figure \ref{Fig:cpartaAA} we observe a strong dependence
of this phenomenon on the breaking of parity, in terms of the ratio
$A_1/A_2$. As discussed in \ref{sym} above, it must vanish for $A_1 =
A_2$. At the same time, we expect it to diminish in the opposite
extreme of $A_1 \ll A_2$ or v.v., as in this limit the effect of the
two regions with different field orientation breaking the symmetry
(\ref{rotdiag}) is lost. Therefore there should exist an optimum for
angular momentum separation at intermediate values of $A_1/A_2$.

\section{Conclusion}\label{conc}
With this work we intend to demonstrate the feasibility of pumping
angular momentum in a chaotic scattering system, in the regime of fast
and strong driving where adiabatic or perturbative methods would
fail. The transport phenomena we observe, both of mass and angular
momentum, owe themselves to an interplay of strongly nonlinear
dynamics and the breaking of spatio-temporal symmetries. They go far
beyond the frame of linear response and do not require a two-parameter
driving as in adiabatic pumping. As a welcome side effect, the
sensitive parameter dependence of the scattering process opens the
possibility not only to generate angular-momentum currents but also to
control them on short time scales and thus to involve them in
computation processes.

At the same time, our results constitute an important example of how
an internal degree of freedom not only participates in chaotic
scattering, but even gives rise to new phenomena. While in this work
we focus on transport, the aspect of chaotic scattering with internal
freedoms deserves a closer scrutiny in a separate effort.

Another relevant side issue is the fact that spin separation clearly
corresponds to a reduction of entropy. In this paper we remain within
a perfectly deterministic Hamiltonian setup, a more realistic model
including thermodynamic aspects however would urgently have to address
this question.

Similarly, connecting the pump on both sides with reservoirs as is
usually considered in the context of electronic transport phenomena
opens additional possibilities to achieve directed currents. The
distributions characterizing the reservoirs (e.g., Fermi distributions
parameterized by the chemical potential) require dissipative
processes to remain invariant and thus implicitly constitute a time
arrow. As a consequence, in such a configuration directed transport can
arise even if the scattering system itself does not break
time-reversal invariance.

The classical study presented here can be considered as a
preparatory survey aiming towards chaotic spin pumping. On grounds of
semiclassical arguments, we would expect the mechanisms we elucidate
to carry over at least to a ``mildly'' quantum regime. The electron
spin, however, corresponds to the deep quantum limit where among other
consequences, the assumption of a slow spin precession, comparable
to the other time scales in the system, no longer applies. In order to
address this question, we depict in figure \ref{Fig:fastpres}b mass and
spin currents as a function of the magnetic field strength, equivalent
to the precession frequency, at constant asymmetry. We see that
directed transport does not diminish significantly and we even find
instances of pure spin current in a regime where both time scales
differ by more than an order of magnitude, giving us some confidence
that chaotic pumping of angular momentum could extend down to $s =
\hbar/2$.

\ack
We enjoyed fruitful discussions with F Leyvraz, J Mahecha, K Richter,
and C Viviescas. Financial support by Volkswagen Foundation (grant
I/78235), Colciencias (grant 1101-05-17608), and Universidad Nacional
de Colombia (grant DIB-803940) is gratefully acknowledged. One of us
(FLD) thanks for a PhD studentship by Universidad Nacional de Colombia
in the program \emph{Becas para Estudiantes Sobresalientes de
Posgrado}.

\end{document}